\documentclass[10pt,aip,apl,twocolumn,showpacs,showkeys]{revtex4-1}

\usepackage{graphicx}% Include figure files
\usepackage{dcolumn}% Align table columns on decimal point
\usepackage{bm}
\usepackage{color}
\begin{document}

\title{Spin Hall Effect-driven Spin Torque in Magnetic Texture}
\date{\today}
\author{A. Manchon$^{a)}$}
\affiliation{Division of Physical Science and Engineering, KAUST, Thuwal 23955, Saudi Arabia.}
\author{K.-J. Lee$^{b)}$}
\affiliation{Department of Materials Science and Engineering, Korea University, Seoul 136-713, Korea.}

\begin{abstract}
Current-induced spin torque and magnetization dynamics in the presence of spin Hall effect in magnetic textures is studied theoretically. The local deviation of the charge current gives rise to a current-induced spin torque of the form $(1-\beta{\bf M})\times[({\bf u}_0+\alpha_H{\bf u}_0\times{\bf M})\cdot{\bm \nabla}]{\bf M}$, where ${\bf u}_0$ is the direction of the injected current, $\alpha_H$ is the Hall angle and $\beta$ is the non-adiabaticity parameter due to spin relaxation. Since $\alpha_H$ and $\beta$ can have a comparable order of magnitude, we show that this torque can significantly modify the current-induced dynamics of both transverse and vortex walls.\\
\begin{small}
$^{a)}$aurelien.manchon@kaust.edu.sa\\
$^{b)}$kj\_lee@korea.ac.kr
\end{small}
\end{abstract}
\maketitle
The study of the interplay between magnetization dynamics and spin-polarized currents through Spin Transfer Torque \cite{stt,zhangli,thiaville,barnes1,tatara} (STT) has culminated with the observation of current-induced domain wall motion and vortex oscillations, revealing tremendously rich physics and dynamical behaviors \cite{beta1,beta2,beta3}. The precise nature of STT in domain walls is currently the object of numerous investigations both experimentally and theoretically. One of the important issues is the actual magnitude of the so-called {\em non-adiabatic} component of the spin torque $\beta$ \cite{beta1,beta2,beta3,seo}.\par

In addition, the nature of spin torque in the presence of spin-orbit coupling (SOC) has been recently uncovered. Although SOC has been long known to generate magnetization damping and spin relaxation, recent studies have suggested that specific forms of {\em structure-induced} spin-orbit coupling could act as a source for the spin torque \cite{manchon1}. However, in the case of {\em impurity-induced} SOC, incoherent scattering averages out the spin accumulation so that no SOC-induced spin torque can be generated in homogeneous ferromagnets \cite{manchon2}. Nevertheless, SOC-induced asymmetric spin scattering by impurities in ferromagnetic materials generates Anomalous Hall Effect (AHE), creating a charge current transverse to both the injected electron direction and the local magnetization \cite{sinovareview}. Interestingly the Hall angle $\alpha_H$, defined as the amount of deviated charge current, can be as large as a few percent in thin films \cite{sinovareview}, which is on the same order of magnitude as the non-adiabatic coefficient $\beta$ \cite{beta1,beta2,beta3}. Therefore, it seems reasonable to wonder whether anomalous charge currents could have a sizable effect on domain wall velocities.\par

In this letter we study the influence of such a transverse charge current on current-induced domain wall motion. We show that this AHE-induced charge current generates an additional torque component, proportional to the Hall angle, along the direction perpendicular to both the charge injection direction ${\bf u}_0$ and to the local magnetization ${\bf M}$ ($\propto\alpha_H[({\bf u}_0\times{\bf M})\cdot{\bm \nabla}]{\bf M}$). The current-driven magnetization dynamics in transverse and vortex walls is analyzed using Thiele formalism.\par

The mechanisms underlying AHE have been studied experimentally and theoretically for more than 60 years (see Ref. \onlinecite{sinovareview} for a comprehensive review). For the transport regime we are interested in (good metal regime, with a conductivity $\approx 10^4-10^6\Omega^{-1}$cm$^{-1}$), the transport is dominated by scattering-independent mechanisms, i.e. intrinsic/side-jump contributions \cite{sinovareview}.
% In a recent comprehensive study by Tian et al. \cite{tian}, the authors showed that epitaxially grown Fe layers display such scattering-independent anomalous conductivity for thicknesses larger than 10 nm. In the following, we will limit our study to weakly disordered magnetic layers with thickness larger than 10nm. 
Disregarding the effect of band structure-induced SOC, we will treat the spin transport within the first order Born approximation, only accounting for the anomalous velocity arising from side-jump scattering \cite{lyo,zhang}.\par

We adopt the conventional {\em s-d} Hamiltonian, where the electrons responsible for the magnetization and the ones responsible for the current are treated separately and coupled through an exchange constant $J$. We also take into account an impurity potential $V_{imp}$ and its corresponding spin-orbit coupling acting on the itinerant electrons. The one-electron Hamiltonian reads
\begin{eqnarray}\label{eq:H}
{\hat H}&=&\frac{{\hat{\bf p}}^2}{2m}+J{\hat{ \bm \sigma}}\cdot{\bf M}+\frac{\xi}{m}\left({\hat{\bm \sigma}}\times{\bm \nabla}V_{imp}\right)\cdot{\hat{\bf p}}+V_{imp},
\end{eqnarray}
In Eq. (\ref{eq:H}), the hat $\hat{}$ denotes an operator while the bold character indicates a vector. ${\hat{\bm \sigma}}$ is the vector of Pauli spin matrices, $\xi$ is the spin-orbit strength (as an estimation, $\xi\approx10^{-17}-10^{-19}$s) and $V_{imp}=V_{imp}({\hat{\bf r}})$ is the impurity potential which is spin-dependent ($2\times2$ matrix) in principle. The magnetization direction ${\bf M}({\bf r},t)=(\sin\theta\cos\phi,\sin\theta\sin\phi,\cos\theta)$ varies slowly in time and space, so that the itinerant electron spins closely follow the magnetization direction (adiabatic approximation). In this picture, the velocity operator is \cite{lyo}
\begin{eqnarray}\label{eq:v}
{\hat {\bf v}}=-i\frac{\hbar}{m}{\bm \nabla}+\frac{\xi}{m}{\hat{\bm \sigma}}\times{\bm \nabla}V_{imp}.
\end{eqnarray}
%Although more complete approaches allow for the description of intrinsic, skew scattering and side-jump sources of anomalous Hall effect \cite{sinovareview}, the present method offers the advantage of providing a simple framework for a reasonable qualitative analysis of the torque governed by spin-orbit coupling in spin texture.
The expectation value of the velocity in the presence of spin-orbit coupling has been worked out by several authors \cite{lyo,zhang} and can be written
\begin{eqnarray}\label{eq:vel}
\langle{\bf{\hat v}}\rangle&=&\frac{1}{i\hbar}\langle[{\bf{\hat r}},H]\rangle\approx{\bf v}+\xi {\hat T}{\bf v}\times{\bm{\hat \sigma}},\\
&&{\hat T}=\Sigma_\tau{\hat I}+\Delta_\tau{\bm{\hat \sigma}}\cdot{\bf M}.
\end{eqnarray}
Here, $\Sigma_\tau=1/\tau^\uparrow+1/\tau^\downarrow$, $\Delta_\tau=1/\tau^\uparrow-1/\tau^\downarrow$, $\tau^\sigma$ being the spin-dependent electron momentum relaxation time. The form of the anomalous velocity displayed in Eq. (\ref{eq:vel}) is the extension of the anomalous velocity derived in Ref. \onlinecite{zhang} to non-collinear magnetic textures. 

The description of diffusive non-collinear spin transport in ferromagnets has been intensively addressed over the past ten years \cite{brataas, zlf, barnas} using different approaches. As an example, for slowly varying magnetization in the presence of anomalous velocity, Eq. (\ref{eq:v}), the relaxation time approximation of the Boltzmann formalism yields a spinor current of the form \cite{zlf}
\begin{eqnarray}
\underline{\underline{{\cal J}}}&=&{\hat C}({\bf {\hat E}}-{\bm\nabla}{\hat \mu}({\bf r}))+{\hat C}_H({\bf {\hat E}}-{\bm\nabla}{\hat \mu}({\bf r}))\times{\bf M},
\end{eqnarray}
where ${\hat C}=\frac{1}{2}C_0({\hat I}+P{\bf M}\cdot{\bm \sigma})$ is the normal conductance, ${\hat C}_H=\frac{1}{2}C_{H0}({\hat I}+P_H{\bf M}\cdot{\bm \sigma})$ is the anomalous conductance and ${\hat \mu}({\bf r})$ is the local spin-dependent electro-chemical potential. ${\bf E}$ is the electric field and $P$ ($P_H$) is the polarization of the (anomalous) conductivity. This form is very similar to the one derived by Zhang \cite{zhang}, extended to non collinear magnetization situations. The charge and spin currents are calculated using the spinor definition
\begin{eqnarray}\label{eq:1}
{\bf J}_e=Tr[\underline{\underline{{\cal J}}}{\hat I}],\;{\cal J}_s=-\frac{\mu_B}{e}Tr[\underline{\underline{{\cal J}}}{\hat {\bm \sigma}}].
\end{eqnarray}

In addition, the spin density continuity equation can be extracted from Eq. (\ref{eq:H}) using Ehrenfest's theorem and in the lowest order in SOC
\begin{eqnarray}\label{eq:st}
&&\partial_t{\bf m}=-{\bf \nabla}\cdot{\cal J}_s-\frac{1}{\tau_{J}}\delta{\bf m}\times{\bf M}-\frac{\delta{\bf m}}{\tau_{sf}},
\end{eqnarray}
Where ${\bf m}=n_0{\bf M}+\delta{\bf m}$ ($n_0$ is the equilibrium itinerant spin density) and the spin current is defined as: ${\cal J}_s=\langle\langle{\bm {\hat \sigma}}\otimes{\bf{\hat v}}+{\bf{\hat v}}\otimes{\hat \sigma}\rangle\rangle_{\bf{k}}$, ${\bf{\hat v}}$ being the velocity operator defined in Eq. (\ref{eq:v}). The inner brackets $\langle...\rangle$ denote quantum mechanical averaging and the outer brackets $\langle...\rangle_{\bf k}$ refer to ${\bf k}$-state average, $\otimes$ being the direct product. By simply injecting the spin current Eq. (\ref{eq:1}) into Eq. (\ref{eq:st}), we obtain the explicit spin continuity equation
\begin{eqnarray}\label{eq:spin1}
\partial_t {\bf m}&=&\frac{\mu_B}{e}[(P C_0{\bf E}+P_HC_{H0}{\bf E}\times{\bf M})\cdot{\bm \nabla}]{\bf M}\nonumber\\
&&-\frac{1}{\tau_{J}}\delta{\bf m}\times{\bf M}-\frac{\delta{\bf m}}{\tau_{sf}}.
\end{eqnarray}

To obtain a tractable form of the spin torque, we assume $P\approx P_H$ and $\alpha_H=C_{H0}/C_0$. After manipulating Eq. (\ref{eq:spin1}) (see Ref. \onlinecite{zhangli}), the spin torque ${\bf T}=\frac{1}{\tau_{J}}\delta{\bf m}\times{\bf M}$ in adiabatic approximation reads 
\begin{eqnarray}\label{eq:st2}
{\bf T}&=&(1-\beta{\bf M}\times)\left[-n_0\partial_t{\bf M}+b_J[{\bf u}\cdot{\bm\nabla}]{\bf M}\right],
\end{eqnarray}
where $b_J=\mu_B P C_0E/e$, ${\bf u}={\bf u}_0+\alpha_H{\bf u}_0\times{\bf M}$, ${\bf u}_0=C_0{\bf E}/|C_0{\bf E}|$ being the injected current direction. One recognizes the renormalization torque ($\propto \partial_t {\bf M}$), the usual adiabatic and non-adiabatic spin torque \cite{zhangli,thiaville} ($\propto b_J$ and $\beta b_J$) and the AHE-induced torques ($\propto \alpha_H b_J$ and $\propto \beta\alpha_H b_J$). Note that recently, Shibata and Kohno have derived a similar form for the spin torque in magnetic texture in the case of skew scattering \cite{shibata2}. In the case of slowly varying magnetic texture ($\partial_t {\bf M}\rightarrow 0$), the spin torque becomes
\begin{eqnarray}\label{eq:st3}
{\bf T}&=&b_J(1-\beta{\bf M}\times)[({\bf u}_0+\alpha_H{\bf u}_0\times{\bf M})\cdot{\bm\nabla}]{\bf M}.
\end{eqnarray}

To extract the dynamics induced by these additional terms, we analyze the current-driven domain wall motion using Thiele free energy formalism \cite{thiele} for rigid domain wall motion ($\partial_t{\bf M}=(-{\bf v}\cdot{\bm \nabla}){\bf M}$). Thiele's dynamic equation yields

\begin{eqnarray}\label{eq:stbj2}
&&\int dV\left[{\bf F}+{\bf G}\times({\bf v}+b_J{\bf u})+{\underline{\underline D}}\cdot(\alpha{\bf v}+\beta b_J{\bf u})\right]=0\\
&&{\bf F}={\bm \nabla}W,\;{\bf G}=-\frac{M_s}{\gamma}({\bm\nabla}\theta\times{\bm\nabla}\phi)\sin\theta,\nonumber\\
&&D_{ij}=-\frac{M_s}{\gamma}(\nabla_i\phi\nabla_j\phi\sin^2\theta+\nabla_i\theta\nabla_j\theta).\nonumber
\end{eqnarray}
Here, ${\bf F}$ refers to the external force, ${\bf G}$ the gyrocoupling vector and the ${\underline{\underline D}}$ the dissipation dyadic exerted on the domain wall. The magnetic energy is 
$W=\frac{A}{M_s}({\bm\nabla}{\bf M})^2+\frac{K}{M_s}({\bf M}\times{\bf x})^2+\frac{K_d}{M_s}({\bf M}\times{\bf z})^2-{\bf H}\cdot{\bf M}$, where $A$ is the ferromagnetic exchange, $M_s$ the saturation magnetization, $K$ ($K_d$) is the anisotropy (demagnetizing) energy and ${\bf H}$ is the external field.\par

Let us first consider a magnetic wire along $x$ containing an out-of-plane transverse wall defined by $\theta(x)=2\arctan e^{\frac{sx}{\Delta}},\;\phi=\phi(t)$ ($s=\pm1$). The external force reduces to ${\bf F}=2sH_zM_s/\Delta {\bf z}$, while the gyrocoupling force vanishes (${\bm \nabla}\phi=0$). The final velocity is then:
\begin{equation}\label{eq:vel2}
{\bf v}=\frac{1}{\alpha}(s\gamma H_zM_s-b_J(\beta u_x-\alpha_H\frac{\pi}{4}u_z\sin\phi)).
\end{equation}
Interestingly, the AHE-induced spin torque only acts on the domain wall when injecting the current {\em perpendicular} to the magnetic wire ($u_x=0,\;u_z=1$). Still, in this latter case, the velocity depends on $\sin\phi$ which is in principle time dependent. This quantity can be determined through the Landau-Lifshitz-Gilbert equation:
\begin{eqnarray}\label{eq:phi}
\partial_t\phi&=&\gamma H_z+\frac{s}{\Delta}(\beta-\alpha)b_J\frac{\pi}{4}\alpha_Hu_z\sin\phi\\
&&-\gamma\alpha H_K\cos\phi\sin\phi.\nonumber\end{eqnarray}

Above Walker breakdown ($\partial_t\phi\neq0,\langle\sin\phi\rangle_t=0$), the velocity, Eq. (\ref{eq:vel2}), does not show any dependence on the perpendicular current. On the other hand, below Walker breakdown ($\partial_t\phi=0$) Eq. (\ref{eq:phi}) provides an implicit expression for the angle $\phi$. In the absence of external field ($H_z=0$)and in the presence of in-plane anisotropy ($H_K\neq0$), $\phi\approx0$ and the velocity is directly proportional to the Hall angle: ${\bf v}\approx\frac{\alpha_H}{\alpha}\frac{\pi}{4}b_J$. Since $\alpha_H$ can be as large as a few percent \cite{sinovareview}, the expected velocity is similar to the once driven by the non-adiabatic spin torque when the current is injected along the structure ($u_x=1,\;u_z=0$).\par

In the absence of in-plane anisotropy field ($H_K=0$), the domain wall velocity becomes independent of the current density and reduces to $v=\frac{s}{\alpha}\gamma\Delta H_z\frac{\alpha-2\beta}{\alpha-\beta}$. This indicates that the anomalous current only distorts the domain wall structure, without inducing any displacement. \par

In contrast with transverse walls, vortex walls present a 2-dimensional texture that couples longitudinal and transverse current-induced velocities \cite{he,moon}. The vortex wall is located at the center of a magnetic layer and in the vortex region, the angles are:
\begin{eqnarray}\label{eq:stbj4}
&&\theta_{{\bf r}^2<r_0^2}=2s\tan^{-1}\frac{r}{r_0},\;\theta_{R^2>{\bf r}^2>r_0^2}=\pi/2,\\
&&\sin\phi_{{\bf r}^2<R^2}=c\frac{x}{r},\;\cos\phi_{{\bf r}^2<R^2}=-c\frac{y}{r},
\end{eqnarray}
where ${\bf r}=x{\bf e}_x+y{\bf e}_y$, $c$ ($s$) is the chirality (polarity) of the vortex. In principle, the vortex extends up to a radius R, beyond which the domain wall can be modeled as transverse walls. In the presence case, we do not consider the action of these outer transverse walls and concentrate on the vortex wall core dynamics. From Thiele free energy, Eq. (\ref{eq:stbj2}), we get the coupled equations for longitudinal and transverse velocities:
\begin{eqnarray}
\alpha {\cal C}v_x-v_y&=&-[\beta{\cal C}+\frac{\alpha_H}{2}]b_Ju_x,\\
v_x+\alpha{\cal C}v_y&=&-[1-\beta\frac{\alpha_H}{2}{\cal D}]b_Ju_x,
\end{eqnarray}
where ${\cal C}=1+\frac{1}{2}\ln\rho$, ${\cal D}=1+\ln\frac{2\rho}{1+\rho^2}$ and $\rho=R/r_0$. This yields the velocities \cite{moon}:
\begin{eqnarray}
v_x&=&-\frac{1+\alpha\beta{\cal C}^2+\frac{\alpha_H}{2}(\alpha{\cal C}-\beta{\cal D})}{1+\alpha^2{\cal C}^2}b_Ju_x,\\
v_y&=&\frac{(\beta-\alpha){\cal C}+\frac{\alpha_H}{2}(1+\alpha\beta{\cal CD})}{1+\alpha^2{\cal C}^2}b_Ju_x.
\end{eqnarray}
It clearly appears that AHE significantly influences the motion of a vortex core by enhancing the transverse velocity $v_y$. As an illustration, Fig. 1(a) displays the current-induced polar angle of the core, $\theta=\tan^{-1}v_y/v_x$, as a function of $\rho$, for different values of the Hall angle $\alpha_H$. It indicates that the presence of AHE clearly enhances the polar angle by several degrees.

\begin{figure}
	\centering
		\includegraphics[width=8cm]{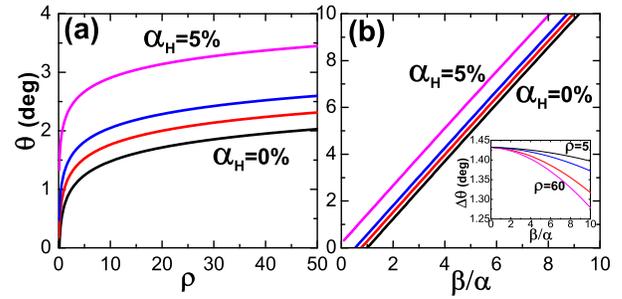}\\
	\caption{\label{fig:Fig1}(Color online) (a) Polar angle as a function of $\rho$ for $\alpha_H=0,0.01,0.02,0.05$; (b) Polar angle as a function of non-adiabaticity $\beta/\alpha$ for $\alpha_H=0,0.01,0.02,0.05$. Inset: $\Delta\theta=\theta(\alpha_H=5\%)-\theta(0)$ as a function of non-adiabaticity for $\rho=5,10,30,60$.}
\end{figure}

Whereas the polar angle is linear as a function of non-adiabaticity [Fig. 1(b)], the influence of AHE-induced torque can be quite significant, especially in the case of sharp vortex core (see Fig. 1(b), inset). These results show that AHE can contribute to more than half of the transverse velocity in the case of current-driven vortex wall motion.\par

In conclusion, we showed that in the presence of SOC, the spin transfer torque exerted on magnetic textures has the general form $(1-\beta{\bf M})\times[({\bf u}_0+\alpha_H{\bf u}_0\times{\bf M})\cdot{\bm \nabla}]{\bf M}$. Whereas the additional AHE-induced torque can induce domain wall motion when injecting the current {\em perpendicular} to a transverse wall, it can also significantly affect the velocity of vortex cores by increasing the {\em transverse} velocity.\par\par
%\section*{Aknowledgement}
K.J.L. acknowledges financial support from NRF grant funded by the Korea government (MEST) (Grant No. 2010-0023798).
\begin{small}

\end{small}
\end{document}